\documentclass{aa}
\usepackage{natbib}
\usepackage{graphicx}
\usepackage[english]{babel}
\usepackage{lscape}

\newcommand{\gppr}{\stackrel{>}{\scriptstyle \sim}}
\newcommand{\gappr}{\raisebox{-0.4ex}{$\gppr$}}
\newcommand{\lppr}{\stackrel{<}{\scriptstyle \sim}}
\newcommand{\lappr}{\raisebox{-0.4ex}{$\lppr$}}

\newcommand{\Porb}{\mbox{$P_\mathrm{orb}$}}
\newcommand{\alc}{\mbox{$\alpha_\mathrm{c}$}}
\newcommand{\alh}{\mbox{$\alpha_\mathrm{h}$}}

\newcommand{\Mwd}{\mbox{$M_\mathrm{wd}$}}
\newcommand{\Msec}{\mbox{$M_\mathrm{sec}$}}
\newcommand{\Rwd}{\mbox{$R_\mathrm{wd}$}}

\newcommand{\Rcrn}{\mbox{$R_\mathrm{crit0}$}}
\newcommand{\Rtid}{\mbox{$R_\mathrm{tid}$}}
\newcommand{\Rout}{\mbox{$R_\mathrm{out}$}}

\newcommand{\Rto}{\mbox{$R_\mathrm{3:1}$}}
\newcommand{\Twd}{\mbox{$T_\mathrm{wd}$}}

\newcommand{\Msun}{\mbox{$M_{\odot}$}}

\newcommand{\Teff}{\mbox{$T_\mathrm{eff}$}}
\newcommand{\Tsec}{\mbox{$T_\mathrm{sec}$}}

\newcommand{\Mtr}{\mbox{$\dot{M}_{\mathrm{tr}}$}}
\newcommand{\Mtrn}{\mbox{$\dot{M}_{\mathrm{tr0}}$}}
\newcommand{\Mcr}{\mbox{$\dot{M}_{\mathrm{cr}}$}}
\newcommand{\Macc}{\mbox{$\dot{M}_{\mathrm{acc}}$}}
\newcommand{\UVcm}{\mbox{$\Delta_{\mathrm{UV,0.5}}$}}
\newcommand{\UVon}{\mbox{$\Delta_{\mathrm{UV,0}}$}}
\newcommand{\EUVcm}{\mbox{$\Delta_{\mathrm{EUV,0.5}}$}}
\newcommand{\EUVon}{\mbox{$\Delta_{\mathrm{EUV,0}}$}}

\begin{document}

\title{Delays in dwarf novae {\rm{\bf II}}: VW\,Hyi, the tidal instability and
  enhanced mass transfer models}
\authorrunning{Schreiber, Hameury, and Lasota}
\titlerunning{Delays in Dwarf Novae {\rm{\bf II}}}
\author{Matthias R. Schreiber\inst{1}, Jean-Marie Hameury\inst{1}, \and Jean-Pierre Lasota\inst{2}}
\institute{ UMR 7550 du CNRS, Observatoire de Strasbourg, 11 rue
de l'Universit\'e, F-67000 Strasbourg, France \\
\email{mschrei@astro.u-strasbg.fr, hameury@astro.u-strasbg.fr}
\and
Institut d'Astrophysique de Paris, 98bis Boulevard Arago, 75014 Paris, France\\
\email{lasota@iap.fr} }
\offprints{M.R. Schreiber}

\date{Received / Accepted }

\abstract{
We discuss the multi--wavelength predictions of the two
models proposed for SU\,UMa stars, i.e the enhanced mass transfer
(EMT) and the tidal thermal instability (TTI) models. We
focus on the systematic differences of the suggested scenarios
before discussing the model predictions together with the
observations of the best-studied SU\,UMa system, VW\,Hyi. We find
that assuming the standard form of the viscosity parameter
$\alpha$, both models predict only outbursts being triggered at the inner
edge of the accretion disc. In the
TTI model  the superoutbursts are triggered when the {\em
outer radius} of the disc reaches a certain value, i.e. the 3:1
resonance radius. In contrast, the EMT scenario predicts
superoutbursts when the {\em disc mass} exceeds a critical value.
This causes the EMT model to be much more sensitive to mass
transfer variations than the TTI model. In both models
we find the predicted UV and EUV delays in agreement with the
observations of VW\,Hyi for $\alh/\alc\lappr\,4$. In addition both
models can generate precursor outbursts which are more pronounced
at short wavelengths, in agreement with observations. Variations
found in the observed light curve of single systems (e.g. VW\,Hyi)
as well as the difference between ordinary SU\,UMa stars and
ER\,UMa systems are a natural outcome of the EMT model while the
TTI model fails to explain them.
 }\maketitle

\section{Introduction}

SU\,UMa stars \citep[e.g.][for a review]{warner95-1} are
short-period, i.e. $\Porb\leq\,2.2$\,hr\footnote{CVcat
\citep{kubeetal03-1} lists five exceptions: MN\,Dra
($\Porb=2.5$\,hr), NY\,Ser ($2.7$\,hr) ,TU Men ($2.8$\,hr), V405
Vul ($2.9$\,hr), and ES\,Dra ($4.2$\,hr).} dwarf novae whose light
curve consists of two types of outburst: normal dwarf nova
outbursts and 5-10 times longer as well as $\sim0.7$\,mag brighter
superoutbursts. SU\,UMa stars can be divided in three subgroups:
ordinary SU\,UMa stars show several normal outbursts between two
superoutbursts and the whole supercycle proceeds on a timescale of
several $10^2$\,d; WZ Sge systems are SU\,UMa stars with extremely
long recurrence times of superoutbursts (several $10^3$ d) and no
normal outbursts in between; finally ER\,UMa stars are defined by
their very short supercycles ($\sim\,50$\,d) and frequent normal
outbursts. In fact ER\,UMa systems are hardly ever in a
quiescent state.

Superoutbursts differ from the normal ones not only in their
length and amplitude but also in pronounced periodic humps found
in the orbital light curve. The period of these so--called
superhumps is a few percent longer than the orbital one and the
phenomenon is generally thought to result from disc deformations
when the radius of the disc reaches the 3:1 resonance radius
\citep{whitehurst88-1,whitehurst+king91-1,hirose+osaki90-1,lubow91-1}.
The resonance is possible only if the mass ratio of the components
is small, i.e. $q\equiv\Msec/\Mwd\lappr\,0.33$. Often superhumps
are considered as a defining feature of superoutbursts but we
advocate caution with this definition because e.g. the dwarf nova
U\,Gem showed (during more than 100 years of observations) a
single superoutburst without a superhump \citep[see][for a
detailed discussion]{lasota01-1,smak00-1}.

Normal outbursts of SU\,UMa stars are thought to be the usual
dwarf-nova outbursts described by the disc instability model
(DIM) \citep[see][for a recent review]{lasota01-1}. This model is
based on the existence of a thermal-viscous instability in regions
where hydrogen is partially ionized, and the opacities depend
strongly on temperature. If one plots the effective disc
temperature $T_{\rm eff}$ at a given radius $r$ (or equivalently
the mass transfer rate $\dot{M}$) as a function of the disc
surface density $\Sigma$, one obtains the well known S-curve, in
which the upper and lower branches are stable and the intermediate
one is unstable. These branches are delimited by two critical
values of $\Sigma$, $\Sigma_{\rm max}$ above which no cool
solution exists, and $\Sigma_{\rm min}$ below which no hot
solution is possible.

Although there is general agreement that the mechanism causing
normal outbursts is the thermal-viscous instability and that
superhumps are due to the 3:1 resonance, whether superoutbursts
are also caused by the 3:1 resonance or if conversely
superoutbursts trigger the superhump phenomenon remains an
unanswered and intensively discussed question \citep[see
e.g.][]{smak95-1}. In the DIM framework, superoutbursts occur if
sufficient mass is accumulated in the supercycle and if during the
(super)outburst itself the onset of the cooling front can be
sufficiently delayed. Currently two scenarios for SU\,UMa
superoutbursts are debated:

\noindent {\bf 1.} \citet{osaki89-1} generalized the DIM  by
adding a ``tidal instability'' thereby developing the thermal
tidal instability model (TTIM) \citep[see][for a
review]{osaki96-1}. In this model an enhanced tidal torque is
assumed to arise when the outer radius of the disc reaches
the 3:1 resonance radius. The additional torque leads to enhanced
tidal dissipation which prevents the onset of a cooling
front. After superoutbursts, the disc is very small and
contains not much mass. Therefore,
before the next superoutburst starts,
the disc goes through a cycle
of several small outbursts
during which its outer edge does
not reach the 3:1 resonance radius and only a very small
fraction of the disc mass is accreted onto the white dwarf.
In the TTI scenario, the superoutbursts are caused by the 3:1 resonance.

\noindent {\bf 2.} An alternative scenario has been put forward
by \citet{vogt83-2}, \citet{smak84-1}, and \citet{osaki85-1}.
These authors suggested that superoutbursts are caused by enhanced
mass transfer (EMT) from the secondary. Indeed,
\citet{hameuryetal00-1} showed that relating the mass transfer
rate to the accretion rate, i.e. assuming that irradiation of
secondary is somehow increasing the mass transfer rate, allows to
reproduce the observed visual light curves. In the EMT
model (EMTM) the high accretion rates during superoutburst are
expected to force the disc to expand beyond the 3:1 resonance
radius thereby accounting for superhumps. In the EMTM
superoutbursts are caused by the enhanced mass transfer and
only superhumps are related to the 3:1 resonance.

It appears difficult (if not impossible) to decide which model
should be preferred based on the analysis of the optical
light curves only: both models introduce a new so far
rather unconstrained parameter (defining either the strength of
the tidal instability or enhanced mass transfer) and both models
are more or less successful in reproducing the light curve as well
as the superhump phenomenon. In addition, the claimed
observational evidence for enhanced mass transfer during outburst
\citep{vogt83-2,smak91-1,smak95-1,pattersonetal02-1} has recently
been questioned by \citet{osaki+meyer03-1}.

Additional information which may help us to constrain the models
comes from simultaneous multi-wavelength observations. In a first
paper \citep[][hereafter paper\,\rm{I}]{schreiberetal03-1} we
analyzed the predictions of the DIM concerning the time lags
between the rise to outburst at different wavelength observed in
the dwarf nova SS\,Cyg. In this paper we use the same version of
the DIM and approximation of the boundary layer to investigate the
multi-wavelength predictions of the two competing SU\,UMa
scenarios by including into the model either EMT or a version of the TTI.
Finally we discuss our numerical results in the light of the observationally
best studied SU\,UMa system, i.e. VW\,Hyi. The structure of the
paper is as follows: we start with a careful review of the
observational constraints available (Sect.\,\ref{s-obs}) before
presenting the predictions of our simulations
(Sect.\,\ref{s-sim}). In the last two sections we discuss our
results in the context of the observations
(Sect.\,\ref{s-sim_obs}) and earlier findings
(Sect.\,\ref{s-prev}).

\section{Reviewing the observations \label{s-obs}}

\begin{figure}
\includegraphics[width=8.5cm, angle=0]{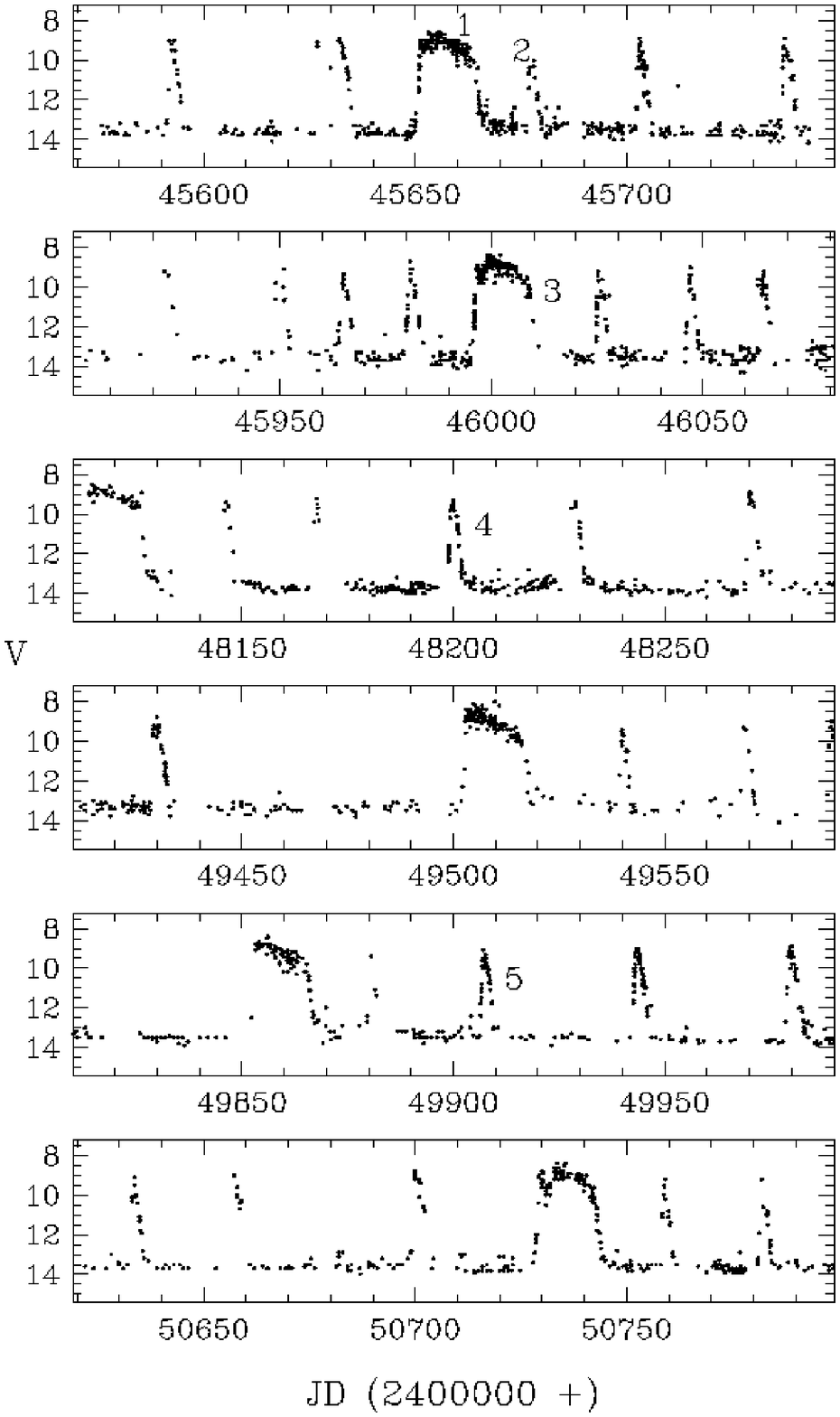}
\caption{\label{f-obs} Snapshots of the visual light curve of
VW\,Hyi. The five panels show parts of the light curve including
the outbursts which additionally have been observed at UV and/or
EUV wavelengths (see Table\,\ref{t-obs}). The fourth panel from
the top shows the optical superoutburst for which
\citet{maucheetal01-1} detected an EUV precursor. The bottom panel
gives an example for a superoutburst with optical precursor.
The data were provided by F. Bateson (VSS\,RASNZ) and J. Mattei (AAVSO).
}
\end{figure}

The long term light curves of SU\,UMa dwarf novae show outburst
cycles consisting of several short outbursts followed by a long
and relatively bright superoutburst.
The phenomenology of the three SU\,UMa subclasses has been
outlined in the introduction. In this section we briefly review
multi-wavelength observations of SU\,UMa systems with particular
emphasis on VW\,Hyi.

\subsection{Visual long-term light curve \label{s-vis}}

\begin{table*}
\begin{center}
\newcounter{ref}
\newcommand{\tcite}{\stepcounter{ref}\arabic{ref}}
\newcommand{\tref}[1]{\stepcounter{ref}(\arabic{ref})\,\citealt{#1}}
\caption[]{\label{t-obs} Outbursts of VW\,Hyi observed at
different wavelength and the resulting delays. The time-resolution
of the Exosat observations (No.\,1) was very poor and the listed
delays are highly uncertain. For outburst No.\,3 a UV precursor
has been detected with Voyager \citep{polidan+holberg87-1}. As in
paper\,I \UVon and \EUVon denote the delays measured at the onset
of the outburst whereas \UVcm and \EUVcm have been determined at
half the maximum optical flux.
The values for the delays are
given in days.
Super and normal outbursts are
labelled with ``s'' respectively ``n''.}
\setlength{\tabcolsep}{2.5ex}
\begin{tabular}{lccccccccr}
\hline\hline\noalign{\smallskip} \multicolumn{1}{c}{No.} &
\multicolumn{1}{c}{Instrument} & JD+2440000 &
\multicolumn{1}{c}{$\Delta_{\mathrm{EUV,0}}$} &
\multicolumn{1}{c}{$\Delta_{\mathrm{EUV,0.5}}$} &
\multicolumn{1}{c}{$\Delta_{\mathrm{UV,0}}$} &
\multicolumn{1}{c}{$\Delta_{\mathrm{UV,0.5}}$} &
\multicolumn{1}{c}{Type} &
Ref.\\
\noalign{\smallskip}
\hline\noalign{\smallskip}
1 & Exosat  &  5650 & $\sim2.5$ & $\sim2.5$  & -- & --    & s & \tcite\\
2 & Voyager  &  5677 & -- & -- & $\sim0.3$ & $\sim0.4$  & n & \tcite\\
3 & Voyager  &  5996 & -- & -- & $\sim0.4$ & $\sim0.5$   & s & 2 \\
4 & IUE      &  8199 & -- & -- & $\sim0.4$ & $\sim0.4$  & n & \tcite\\
5 & EUVE\&Voyager &  9907 & $\sim0.75$ & $\sim0.5$ & $\sim0.5$ &
$\sim0.3$  & n & \tcite\\
\noalign{\smallskip}\hline\noalign{\smallskip}
\end{tabular}
\linebreak
\setcounter{ref}{0}
References:
\tref{vanderwoerd+heise87-1}
\tref{polidan+holberg87-1}
\tref{wheatleyetal96-1}
\tref{mauche02-1}
\end{center}
\end{table*}

Fig.\,\ref{f-obs} shows parts of the long term light curve of
VW\,Hyi. While the duration of normal outburst is around
$3-5$\,days, superoutbursts last essentially longer
($\sim\,10-15\,$\,days\footnote{As in \citet{mohanty+schlegel95-1}
the length has been measured at $V=10$ }). The single outbursts as
well as the outburst cycles are not strictly periodic.
\citet{bateson77-1} classified the superoutbursts of VW\,Hyi into
basically two types: S1-S5 with a single superoutburst (hereafter type\,1) 
and S6-S8 where a precursor outburst is separated from the
superoutburst (type\,2). In addition the number of normal outbursts
observed between two superoutbursts varies from three to seven and
the supercycle duration is ranging from $\sim\,100$ to $\sim\,250$
days \citep[e.g.][]{bateson77-1, mohanty+schlegel95-1}. Such
variations of the supercycle length and the frequency of normal
outbursts can be considered as typical for ordinary SU\,UMa stars.
Similarly, precursor outbursts have been observed not only in
VW\,Hyi but also in other SU\,UMa stars, e.g. T\,Leo
\citep{howelletal99-1}.

\subsection{The delayed UV and EUV rise \label{s-uvdelay}}

Plotting normalized multi-wavelength light curves of dwarf novae,
one finds a delay between the optical rise and the rise at shorter
wavelengths. For VW\,Hyi the delay of the UV and EUV rise has been
observed for several outbursts. Table\,\ref{t-obs} lists the
obtained time lags. We use the same notation as in paper\,I: the
index ``0.5'' denotes the time lags measured at half the maximum
optical light whereas quantities with index ``0'' have been
measured at the onset of the outbursts. The snapshots of the light
curve we present in Fig.\,\ref{f-obs} contain the outbursts for
which multi-wavelength observations exist. These outbursts are
marked with numbers which correspond to those of column 1 in
Table\,\ref{t-obs}. The fourth and last panel from the top show
the optical superoutburst for which \citet{maucheetal01-1}
detected an EUV precursor and a superoutburst with
clear optical precursor.

\subsection{UV and EUV precursor \label{s-uvprec}}

The above described delays are not the only important
multi--wavelengths phenomenon of SU\,UMa stars. One additionally
finds precursors at short wavelength which are often correlated
with an optical precursor outburst (type 2 superoutburst of
VW\,Hyi). However, the October 1984 superoutburst of VW\,Hyi shows
a UV precursor while the optical emission remains nearly constant.
This UV precursor outburst resembles very much a normal outburst
until the UV rebrightens \citep[][their
Fig.\,3]{polidan+holberg87-1}. At EUV wavelength, a similar
behavior has been measured: during superoutburst, the EUV count
rates of VW\,Hyi, OY\,Car and SW\,UMa are reaching a local minimum
$\sim5$\,d after the optical rise
\citep{maucheetal01-1,burleighetal01-1}.

\subsection{X-rays\label{s-xrays}}
Although a detailed X-ray light curve as for SS\,Cyg
\citep{wheatleyetal03-1} is not available for any SU\,UMa system,
the existing X-ray data of VW\,Hyi contain important information.
Recently \citet{pandeletal03-1} recorded X-ray spectra with XMM
during quiescence and estimate a ``boundary layer'' luminosity
of $8\times\,10^{30}$\,erg\,s$^{-1}$ which corresponds to an
accretion rate of
$5\times\,10^{-12}\Msun\,$yr$^{-1}=3\times10^{14}$gs$^{-1}$, in
agreement with earlier findings \citep{bellonietal91-1}. 
These values have to be considered as lower limits for the real
accretion rate as the X-ray luminosity might not represent
the total luminosity of the accretion flow onto the white
dwarf \citep[see again][]{pandeletal03-1}.

The evolution of the X-ray emission is given by
\citet{hartmannetal99-2} who report six BeppoSAX X-ray
observations covering an outburst cycle. They find that the
X-ray emission significantly decreases during the optical outburst
but remains constant during quiescence.

\subsection{Delayed superhumps\label{s-delsup}}

Since the first detection of superhumps in December 1972
\citep{vogt74-1,warner75-1} many SU\,UMa stars have been
observed using high-speed photometry and -- apart from the U\,Gem
outburst mentioned in the introduction -- superhumps appear in
coincidence with superoutbursts. Superhumps develop from one night
to the other \citep[][]{semeniuk80-1} but appear in general 1--2
days after the superoutburst has reached its optical maximum
\citep[e.g.][]{warner95-1}.

\section{Simulations\label{s-sim}}

\begin{table}
\begin{center}
\caption[]{\label{t-par} Binary parameter of VW\,Hyi and model
parameter. \Rto denotes the 3:1 resonance radius, \Rcrn\, the
outer radius for which the disc becomes circular, and \Rtid\, is
the tidal truncation radius. For the EMTM $f_{\rm ill}$ defines
the strength of enhanced mass transfer and $<\Rout>$ the mean
outer radius which is assumed to be given by the average of $r_1,
r_2$, and $r_{\rm max}$ in Table\,1 of \citet{paczynsky77-1}. }
\setlength{\tabcolsep}{2ex}
\begin{tabular}{lccc}
\hline\hline\noalign{\smallskip}
\Porb/hr       & 1.75& &\\
\Twd          & 20\,000\,K & &\\
\Tsec       & 2750\,K & & \\
$\Mwd/\Msun$ & 0.63 & 0.86$^*$ & \\
$\Msec/\Msun$ & 0.11 & &\\
\Rwd/$10^8$\,cm & 8.4& 6.6$^*$& \\
i/$\,^{\circ}$ & $60$& &\\
d/pc & $65$ & & \\
$R_{\rm tid}$/$10^{10}$\,cm & 2.36 & 2.70$^*$ \\
\hline\noalign{\smallskip}
$R_{3:1}$/$10^{10}$\,cm & 2.14 & 2.33$^*$ & TTIM\\
$\Rcrn$/$10^{10}$\,cm & 1.63 & 1.78$^*$ & TTIM\\
$c_1/c_0$ & 30 & & TTIM \\
\hline\noalign{\smallskip}
$<R_{\mathrm{out}}>/10^{10}$cm & 2.1& 2.29$^*$ & EMTM\\
$f_{\rm ill}$ & 0.3 & 0.25$^*$ & EMTM\\
\noalign{\smallskip}\hline\noalign{\smallskip}
\end{tabular}
\linebreak
\end{center}
 $^*$ Masses of white dwarfs estimated from observations are notoriously
uncertain. For VW\,Hyi \citet{schoembs+vogt81-1}
estimate $\Mwd=0.63$ \Msun but more recent studies indicate a larger value,
i.e. $\Mwd=0.86$ \Msun \citep{sionetal97-1}. This leads to uncertainties
in other parameters as listed above. For the larger mass we assumed a smaller
value of $f_{\rm ill}$ to keep the length of the predicted superoutbursts
similar to the observed ones.
Fortunately, the conclusions of this paper are not affected by these
uncertainties. Except if otherwise stated, we use the parameter according to
$\Mwd=0.63\Msun$. 

\end{table}

\begin{figure*}
\begin{center}
\includegraphics[width=8.5cm, angle=270]{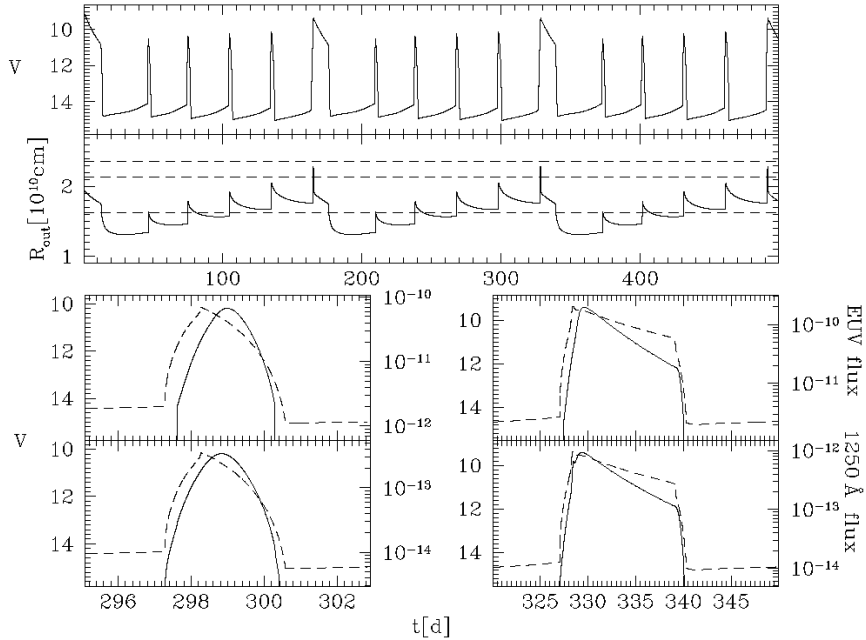}
\caption{\label{f-tti_cls} Calculated light curves using the
standard TTIM. The top panels show the long term evolution of
the visual light curve (V) and the outer radius of the disc
(\Rout). The dashed horizontal lines represent \Rto\, and \Rcrn.
enlargements of the light curve together with the predicted EUV
and UV emission are presented in the lower plots for a normal
outburst (left) and a superoutburst (right). The visual light
curve consists of four small outbursts between two superoutbursts
(top) and the tidal instability ends shortly before the disc
shrinks below \Rcrn (second panel). The bottom panels show the EUV
and UV light curve (solid lines) together with the optical one
(dashed line) for a normal (left) and a superoutburst (right). The
delays close to maximum, i.e. \UVcm and \EUVcm are somewhat
shorter for superoutbursts as the tidal instability is assumed to
develop almost immediately when \Rout reaches \Rto (see text). }
\end{center}
\end{figure*}

We use the DIM-code described in detail in \citet{hameuryetal98-1}
in which tidal dissipation and stream impact heating have been
included \citep{buat-menardetal01-1}.

Recently \citet{smak02-1} analyzed the structure of the outer
regions of accretion discs in dwarf novae and arrived at the
conclusion that neither tidal dissipation nor the stream impact
can play an important role for the disc heating. 
He argues that
(i) most of the impact energy is radiated away by the hot spot so
that not much is left for heating the disc and (ii) the tidal
dissipation affects only a very narrow region of the outer disc
and is (iii) emitted by the edge of the disc. Smak is certainly right in
pointing out that the radiation of the disc's edge should not be neglected.
His arguments against the importance of the stream-impact heating
result from comparing the system luminosity and the hot spot
luminosity of the nova-like system UX\,UMa. Although in principle
correct his study is probably based on a too simple model. In addition, 
Smak's conclusion becomes doubtful due to observational uncertainties. 
As an example we note that even the orbital system parameter for
UX\,UMa derived by \citet{baptistaetal95-1} have recently been
questioned by \citet{froningetal03-1}. Throughout this paper, we
{\em assume} that one half of the stream impact energy is
thermalized in the accretion disc. In any case in our code the
tidal heating is more important. As far as this effect is
concerned Smak's result is based on the calculations of
\citet{ichikawa+osaki94-1} who considered simple periodic particle
orbits in the binary potential. In contrast, 2D-SPH simulations
indicate that a large fraction of the outer disc is strongly
affected by tidal effects \citep{trussetal01-1}. One should also
note that the additional heating of the outer disc seems to be a
crucial ingredient for the DIM to be able to explain dwarf nova
light curves \citep{buat-menardetal01-1} and in particular
the Z Cam phenomenon \citep{buat-menardetal01-2}. We therefore
keep the tidal heating in our calculation, especially as we wish to test the
tidal-thermal instability model which, if Smak were
right, could not be operating in dwarf novae. 

For the emission from the boundary layer we use the simple but
reasonable model presented in paper\,{\rm I}: the boundary
layer is optically thin and emits X-rays for low accretion
rates, but becomes optically thick when the accretion rate exceeds
the critical value $\Mcr=10^{16}$\,gs$^{-1}$. The optically thick
boundary layer emission is approximated by a black body
and assumed to expand with increasing mass accretion rate. It is
this optically thick boundary layer which is the main source of
EUV emission during outburst.
In addition we take into account the
contributions of the (irradiated) secondary (using spectra from
\citet{kurucz93-1}), the hot spot (assuming a black body of
\Teff$=10\,000$\,K), and the white dwarf (black body,
\Twd$=20\,000$\,K).

In quiescence the hot spot emission and the stream-impact heating
contribute significantly to the optical
emission whereas the UV emission is dominated by radiation from the
white dwarf. The effective temperature of the disc is nearly constant around
$\Teff\sim\,3\,000$\,K.
The optically thin boundary layer is assumed to be the only source
of X-ray emission.

During outburst the effective temperature of the disc is following
the classical law ($T_{\rm eff}\propto\,R^{-3/4}$ \citep[see
e.g.][]{lasota01-1}) with effective temperatures of
$\Teff\sim\,10\,000-50\,000$\,K. In this high state the disc is the
main source of optical (outer parts of the disc) and UV (inner
regions) emission. Our approximation of the optically thick
boundary layer (see paper\,I Eqn.\,(1)-(3)) gives a fractional
emitting region of $f\sim2\times10^{-3}$ and a temperature of
$2.5\times10^5$\,K for an accretion rate of
$\Macc=1\times10^{17}$g\,s$^{-1}$. This approximation is probably
not very good as the observed EUV-spectra indicate lower
temperatures and are hardly fitted with a black
body \citep{mauche97-1}. However, as the goal of this paper is to
analyze the chronology of SU\,UMa star outbursts using
different theoretical models (and not to fit the observed EUV
spectra) our conclusions are not affected by this disagreement. 
The binary parameter of VW\,Hyi are given in
Table\,\ref{t-par} and paper {\rm I} contains a more detailed
description of the modeling.

We present the light curves predicted by the model for four
different bands: the optical V band, the 1250\,\AA\, flux density
representing the UV light curve, the integrated 70--130\,\AA\,
flux\footnote{The EUV flux is calculated by integrating the
$70-130$\,\AA\,spectrum using the cross-sections
as a function of wavelength from \citet{morrison+mccammon83-1}.}
assuming $N_H=6\times\,10^{17}$\,cm$^{-2}$, and finally the
accretion rate below $\Mcr$ as a measure for the X-ray emission.
\citep{polidanetal90-1}. The parameters required in the context of
the TTI and the EMT models are also given in Table\,\ref{t-par}.
The time lags predicted by the models between the rise at
different wavelength are compiled in Table\,\ref{t-res}.

We present the calculations of the TTIM as well as the EMTM
and consider the dependence of the predictions on the mass
transfer rate and the inner boundary condition.

\subsection{The TTI model\label{s-sim_tti}}

\begin{figure*}
\begin{center}
\includegraphics[width=8.5cm, angle=270]{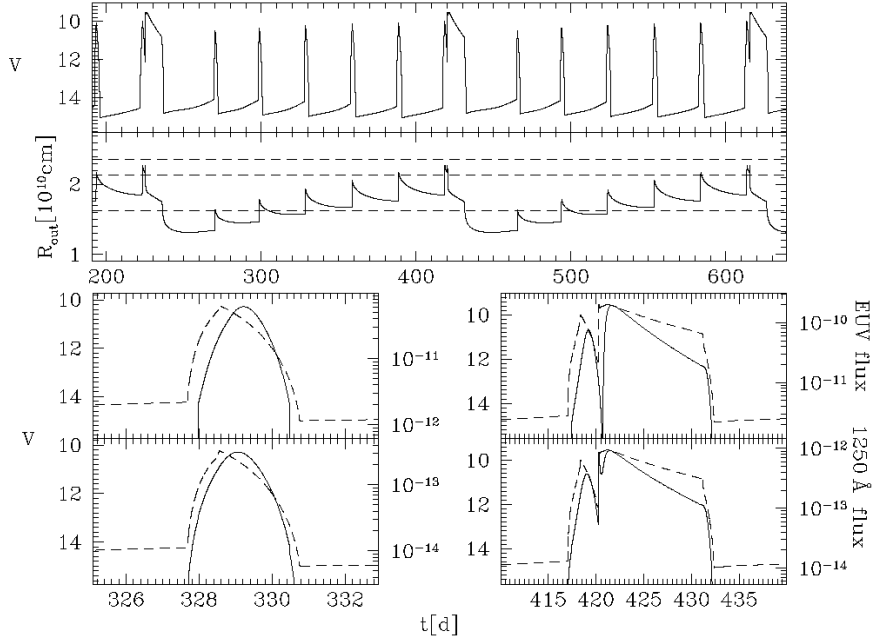}
\caption{\label{f-tti_rev} Calculated light curves assuming a
delay of the tidal instability of 1.75 days. The introduced time
lag leads to a precursor outbursts which is the more pronounced
the shorter the considered wavelength. Left : normal outbursts;
right: superoutbursts (see Fig.\,\ref{f-tti_cls})}
\end{center}
\end{figure*}

Angular momentum conservation in an accretion disc is described
by:
\begin{eqnarray}
j\frac{\partial\Sigma}{\partial\,t}=-\frac{1}{r}
\frac{\partial}{\partial r}(r\Sigma\,jv_{\rm r})&+&\frac{1}{r}
\frac{\partial}{\partial r}
\left(-\frac{3}{2}r^2\Sigma\nu\Omega_{\rm K}\right)\nonumber\\
&+&\frac{j_{\rm 2}}{2\pi\,r}\frac{\partial\dot{M}_{\rm
tr}}{\partial\,r}-\frac{1}{2\pi\,r}T_{\rm tid}(r)
\end{eqnarray}
where $\Sigma$ is the surface density, \Mtr, the mass
transfer rate, $v_{\rm r}$ the radial velocity, $j$ and $j_{\rm 2}$
the specific angular momentum of the disc
and of the
material transferred from the secondary
respectively, and $\nu$ the kinematic
viscosity coefficient \citep[see also][]{hameuryetal98-1}. The
tidal torque ($T_{\rm tid}$) can be approximated by
\begin{equation}
T_{\rm tid}=c\,\Omega_{\rm
orb}\,r\,\nu\,\Sigma\left(\frac{r}{a}\right)^5
\end{equation}
\citep{smak84-1,papaloizou+pringle77-1} where $\Omega_{\rm orb}$
is the angular velocity of the binary orbital motion, $a$ the
binary separation and $c$ a numerical constant taken so as to give
an average disc size equal to a chosen value.

As mentioned in the introduction, accretion discs in dwarf
novae with small mass ratios ($q\lappr0.33$) can expand beyond
the 3:1 resonance radius $\Rto\approx 0.46a$ and become
eccentric. In the framework of the TTIM this is assumed to cause a
strongly increased tidal torque which is usually approximated by
increasing the ``constant'' $c$ by a factor of 20-50. When the
radius of the disc becomes smaller than $\Rcrn=0.35a$, the disc
returns to a circular shape and $c$ decreases to its previous
value. As the tidal instability mainly affects
the outer parts of the accretion disc, we also include a radial
dependence of $c$. When the model predicts the disc to be
eccentric, we use $c(r)=c_1$ for $r>0.9\Rcrn$, $c(r)=c_0$ for
$r<0.8\Rcrn$, and a linear interpolation in between. The tidal
instability will certainly not set in instantly.
We therefore
assume that $c$ varies linearly on short timescales
during the development $t_{\rm d}$ and decay $t_{\rm v}$ of the tidal
instability.
It takes the disc {\em at least} a dynamical
time\footnote{Assuming $r=2\times\,10^{10}$\,cm the dynamical time is
several hundred seconds.} to change its shape. We
assume for the full development of the tidal instability
$t_{\rm d}=2\times10^{4}$\,s.
The time scale of the decay of the tidal instability is at least the time it
takes the disc to dissipate its excess energy.
\citet[][]{buat-menard+hameury02-1} estimated it to
be a few percent of the viscous time in the outer parts of the disc,
i.e. $t_{\rm v}=2\times10^5$\,s which we assume throughout this paper.

\subsubsection{The standard TTIM \label{s-tti_cls}}

Fig.\,\ref{f-tti_cls} shows the light curves at different
wavelength for the standard TTIM. All outbursts are of the
inside--out type i.e. the heating front starts close to the white
dwarf. The cooling fronts of the small outbursts start immediately
after the heating front has reached the outer disc. Between
superoutbursts, the duration and brightness of the small outbursts
slightly increases with time. The final expansion of the disc due
to the tidal instability is simultaneous with the final
optical rise of the superoutbursts.

The UV and the EUV rise are delayed with respect to the
optical rise by 0.2-0.3 and 0.4-0.5\,days respectively. For normal outbursts we
find that both the UV and the EUV delay tend to be $\sim0.1$\,d
longer close to maximum. During superoutbursts, the UV-delay
measured at the beginning of the optical rise does not change but
becomes shorter close to maximum. This is {\bf (1)} because the tidal
instability is assumed to set in immediately and to develop on a
short timescale ($t_{\rm d}=2\times\,10^4$s) when the outer radius
of the disc is reaching the 3:1 resonance radius and {\bf (2)} because
the enhanced tidal dissipation affects simultaneously large parts
of the outer disc ($R>0.8\,\Rcrn$). The almost simultaneous
steepening of the final optical and UV rises of the superoutburst
significantly reduces the predicted UV-delay close to maximum
(\UVcm).

Another general prediction of the standard TTIM which is
worth mentioning is the drastic (i.e. by a factor of
$\gappr\,10$) decrease of the UV and EUV luminosities during
superoutbursts. At the same time the optical emission decreases by
only $\sim\,1.2$ magnitudes. As long as the disc is eccentric, the
enhanced tidal dissipation keeps the disc in the outburst state
although the mass accretion rate decreases significantly. Compared
to the high energy radiation, the optical emission decreases only
slightly as the outer parts of the disc are kept in the high state.
In contrast, the boundary layer and the inner parts of the disc cool
down as the mass accretion rate decreases.

\subsubsection{The timescale of the tidal instability\label{s-tti_rev}}

In the standard TTIM, the time it takes the disc to become
eccentric and the enhanced tidal dissipation to be efficient is
very uncertain. So far we assumed the instability to set in
immediately and to develop rapidly when
$R_{\mathrm{out}}\geq\,R_{\mathrm{3:1}}$. The general coincidence
of superoutburst and superhumps has often been claimed as an
advantage of the TTI model \citep[e.g.][]{osaki96-1}. Almost
as often, the measured 1--2 day delay between the superhump
phenomenon and the superoutburst rise (see Sect.\,\ref{s-delsup})
has been quoted as the principal weakness of the model
\citep[e.g.][]{smak96-1,smak00-1}. Indeed, the sequence of events
predicted by the standard TTIM does not match the observed
one. According to the TTIM, superhumps should appear
simultaneously with the enhanced tidal dissipation, i.e. together
with the final optical rise, but this is not the case (see
Sect.\,\ref{s-delsup}).

In this section we modify the synchronization of the enhanced
tidal dissipation and the observed appearance of superhumps by
including a time lag $t_{\rm l}$ between the moment when the outer
radius reaches \Rto and the onset of enhanced tidal dissipation
due to the instability. Such a revision of the model has been
proposed recently by \citet{osaki+meyer03-1}. We use $t_{\rm
l}=1.75$\,d which accounts for the observed delay between the
optical rise and the appearance of Superhumps (see
Sect.\,\ref{s-delsup}) .

Fig.\,\ref{f-tti_rev} shows the light curves predicted by
this revised TTIM for the relevant wavelengths. Clearly, due
to the revision of the timescale of the tidal instability we
obtain more small outbursts because the outer radius of the disc
needs to be larger than \Rto\, for longer than $t_{\rm l}$ before
the enhanced tidal dissipation sets in. The short time lag
introduced between the expansion of the disc beyond \Rto\, and the
tidal instability leads also to the formation of a cooling
front before the enhanced tidal dissipation can cause an
outside--in heating front and trigger a superoutburst
\footnote{To avoid confusion concerning our previous statement
that all predicted outbursts are of the inside--out type we want
to stress that this is only true if one considers the precursors
being a part of the superoutbursts -- the rebrightening following
the precursor is always of the outside--in type}. Hence, as
predicted by \citet{osaki+meyer03-1}, the introduced time lag
$t_{\rm l}$ forces the model to produce precursor outbursts.
However, Osaki \& Meyer additionally argue that in the revised
scenario the outer radius of the disc may reach the tidal
truncation radius. In contrast, our calculations show that the
disc remains essentially smaller than \Rtid\, even during
superoutburst (Fig.\,\ref{f-tti_rev}) unless the mass transfer rate is very
close to the value required for stable accretion (see next section).
Moreover, if we assume the
parameter corresponding to $\Mwd=0.86$, the difference between \Rtid\, and
\Rto\, increases significantly (see Table \ref{t-par}) and the scenario put
forward by Osaki \& Meyer becomes even more implausible.

The predicted delays of the revised TTI scenario are identical to
those of the standard TTIM which is not at all surprising as
the rise of the normal outbursts and the precursor are not
affected by the revision. Concerning the rebrightening following
the precursor, the predicted UV delay is very short
(\UVcm$\lappr0.05$) while for the EUV delay the value of
\EUVcm$\sim0.5$d still holds. Once the rebrightening is
triggered, the rapidly increasing tidal dissipation affects large
parts of the outer disc and leads to the formation of a fast
heating front which overwhelms the cooling front of the precursor
outburst. As this heating front starts in outer disc regions, the outburst
is of the outside--in type. As analyzed in detail in paper\,\rm{I},
outside--in heating fronts produce shorter delays than inside--out
outburst because the delays depend mainly on the time it takes the
disc to reach high accretion rates which is -- contrary to what
one might naively think -- not at all identical to the time it
takes the heating front to reach the inner edge of the disc. This
important result is also true in the case of the rebrightening.
Moreover, the situation is even more drastic as the tidal
instability is assumed to develop quickly (as superhumps do
\citep{semeniuk80-1}) and to affect simultaneously large parts of
the outer disc. Therefore, the predicted optical and UV emission
rise almost together (as in the standard TTIM picture).
\EUVcm\, depends mainly on the viscous time scale and hence remains
unchanged.
\begin{figure}
\includegraphics[width=8.5cm, angle=0]{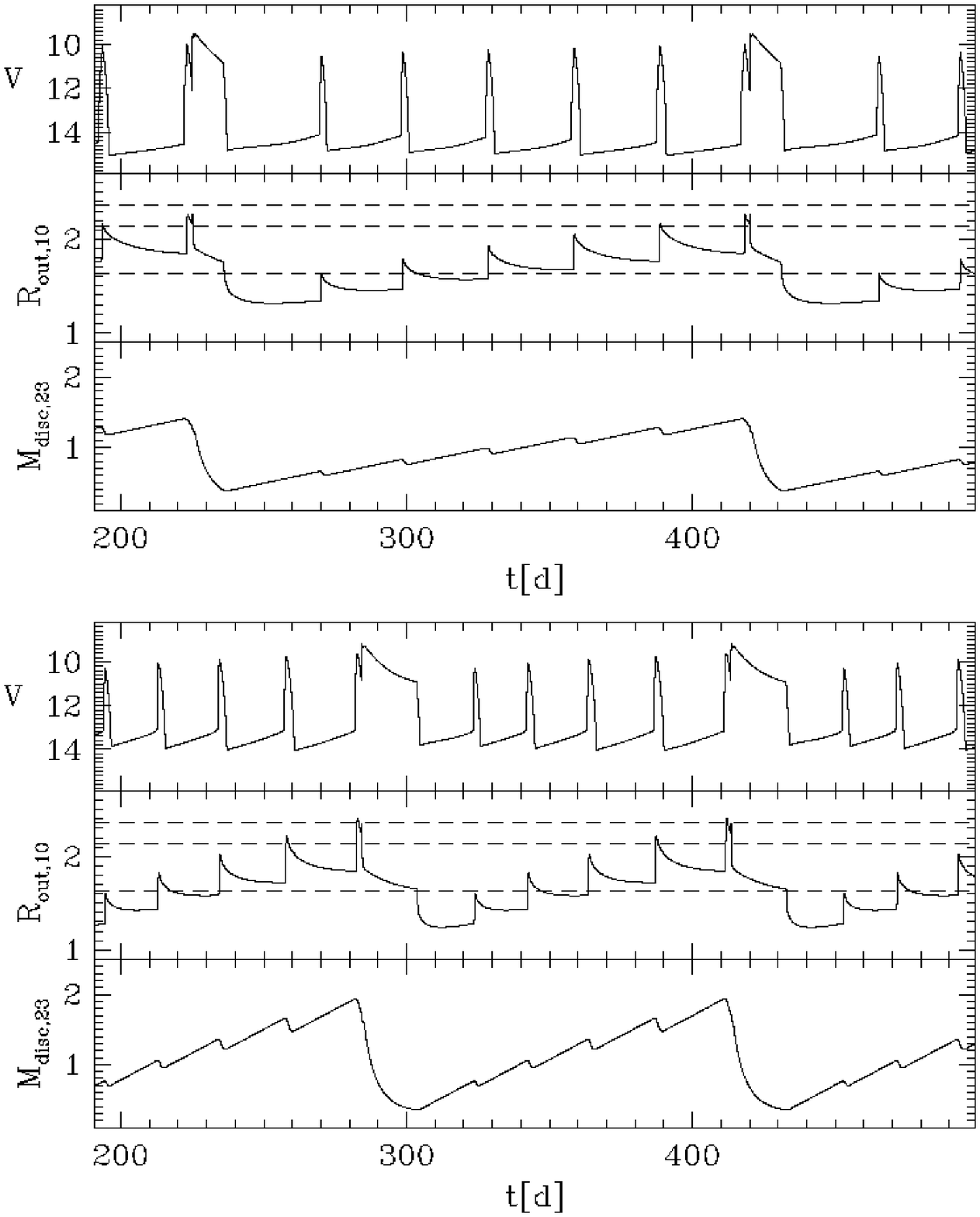}
\caption{\label{f-tti_mtr} Light curves calculated assuming
$t_{\rm l}=1.75$\,d and very different mass transfer rates, i.e.
$10^{16}$\,gs$^{-1}$ (top) and $2.5\times10^{16}$\,gs$^{-1}$
(bottom) . For each value of the mass transfer rate the panels
show the evolution of the visual magnitude, outer radius ($R_{\rm
out,10}=\Rout/10^{10}$\,cm), and the disc mass
($M_{\mathrm{disc,23}}=M_{\mathrm{disc}}/10^{23}$\,g). According
to the TTIM a superoutburst is triggered when the radius of the
disc remains larger than \Rto for $t_{\rm l}$ (see
Sect.\,\ref{s-tti_rev}). The mass of the disc at this time depends
on the mean mass transfer rate.}
\end{figure}

\subsubsection{Dependence on the mass transfer rate \label{s-tti_mtr}}

\begin{figure*}
\begin{center}
\includegraphics[width=9cm, angle=270]{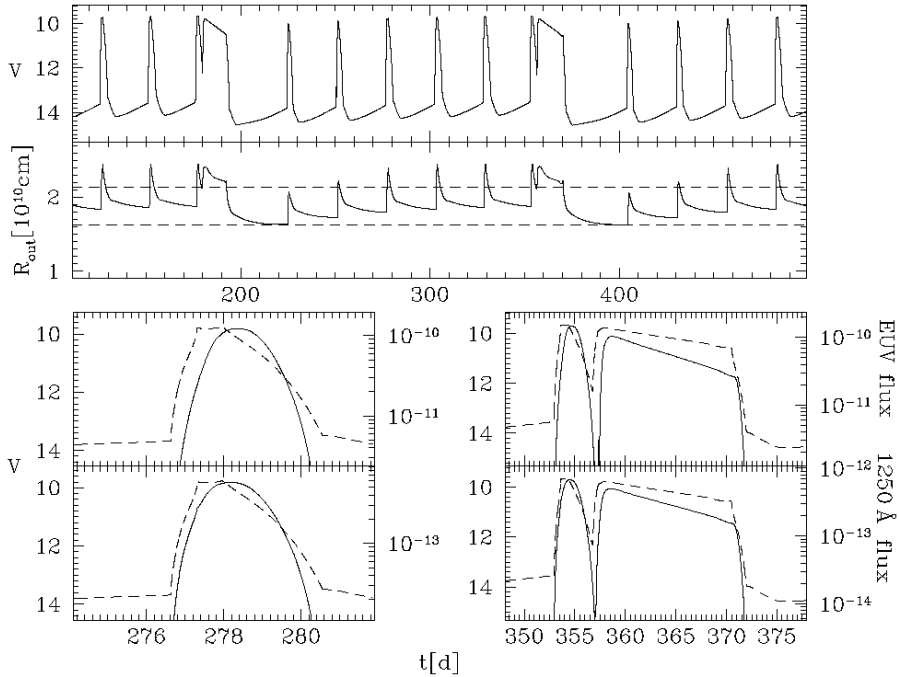}
\caption{\label{f-emt}Multi-wavelength predictions of the EMTM.
Superoutbursts are separated by several small outbursts and the
model predicts precursor outbursts which are more pronounced at
shorter wavelength. The delays measured at the onset of the
optical rise are slightly shorter than the observed ones (compare
Table\,\ref{t-obs} and \ref{t-res}).}
\end{center}
\end{figure*}

As outlined in the introduction and Sect.\,\ref{s-obs}, SU\,UMa
stars show a large variety of features in their light curves
ranging from different supercycle durations to precursor outbursts
at optical and short wavelength. In this context VW\,Hyi is of
vital importance as one finds many of the general SU\,UMa
properties in this particular system: the light curve contains
superoutbursts with and without optical precursor outburst, EUV
and UV precursors have been observed while the optical emission
remained almost constant, and finally the number of normal outbursts
between superoutbursts varies from three to seven. All these
variations should be related to time variations of the mean mass
transfer rate.

We present here the predictions of the revised TTIM for
different mass transfer rates. Fig.\,\ref{f-tti_mtr} shows the
predicted visual light curves assuming $t_{\rm
  l}=1.75$\,d and two values of the mass transfer rate, $\Mtr=1\times\,10^{16}$
and $2.5\times\,10^{16}$gs$^{-1}$. The latter value is very close
to the mass transfer rate above which the disc would be stable.
The diffusion time scale is independent of the mass transfer rate
(see e.g. \citet{lasota01-1}). The recurrence time for inside-out
outbursts should therefore in principle not be affected by
changing the mass transfer rate. However, additional effects like
stream-impact heating (proportional to \Mtr) and tidal dissipation
(proportional to $\Sigma$) which are not included in the
analytical formula of \citet{lasota01-1} are causing a somewhat
reduced recurrence time for significantly increased mass transfer
rate (Fig.\,\ref{f-tti_mtr}). We also obtain longer superoutbursts
and the increased stream-impact heating, tidal dissipation and hot
spot brightness lead to a slightly brighter quiescence. The
general behavior of several small outbursts followed by a
superoutburst, however, does not change even for the drastically
increased mass transfer rate. Superoutbursts are still separated
by several small eruptions. The precursor outburst becomes less
pronounced but always remains visible.

In order to understand the little influence that even very
strong variations of the mass transfer rate have on the predicted
light curve, we have to throw a glance at the triggering condition
for superoutbursts: in the TTIM, a superoutburst starts when
$R_{\rm out}=R_{3:1}$. Increasing the mass transfer rate during
quiescence leads to a shrinkage of the disc. To reach $R_{\rm
out}=R_{3:1}$ the disc therefore needs to contain more mass
(compare the lower panels of Fig.\,\ref{f-tti_mtr}). This
cancels out to some extent the effect of the increased mass
input, and we find the light curve between superoutburst are
rather insensitive to variations of the mean value of the
mass transfer rate. The main effect of the increased mass
transfer rate is that superoutbursts become brighter and longer
due to the increased mass of the disc when such an outburst is
finally triggered. Whether a precursor is visible or not is
also independent of the mean mass transfer rate. It depends only
on the introduced time lag $t_{\rm l}$ if a cooling front starts
before enhanced tidal dissipation sets in.

A final note concerns the evolution of the outer radius of the
disc (second panels of Fig.\,\ref{f-tti_mtr}).
The disc can reach the tidal truncation radius only for mass
transfer rates very close to the value required for stable
accretion and only if the white dwarf mass is not much larger 
than $0.6\,\Msun$.

\subsection{Enhanced mass transfer\label{s-sim_emt}}

As outlined in the introduction, competing with the TTIM is the
enhanced mass transfer scenario. In this picture it is assumed
that the mass transfer rate from the secondary is related to the
flux irradiating the secondary. In our model light curves, we have
already included the contributions of irradiation of the secondary
by the boundary layer and the white dwarf (see paper I). Here we
slightly modify this relation by taking into account the fact that
the accretion luminosity is not entirely radiated away immediately
but that a fraction of it heats the white dwarf and is released
when the white dwarf cools. The cooling of the white dwarf in
VW\,Hyi after outburst has been estimated e.g. by using
UV--spectra during early decline and quiescence
\citep{gaensicke97-1}. The white dwarf returns to its pre-outburst
temperature $\sim\,2$ days after normal and $\lappr\,5$ days after
superoutbursts. We take these results into account by
relating the mass transfer rate to an average (over
$\Delta\,t=2,5$\,days) of the accretion rate. The mass transfer
rate is then determined by
\begin{equation}
\label{eq-emt1}
\Mtr={\rm max}(\Mtr_0, f_{\rm ill} <\Macc>),
\end{equation}
with
\begin{equation}
\label{eq-emt2}
<\Macc>=\int_{-\infty}^{t_0}\Macc\,e^{(-t_0-t)/\Delta\,t}dt,
\end{equation}
\citep[see also][]{buat-menard+hameury02-1}. This is an apparently
arbitrary prescription but -- as we will see -- the general
conclusions of this paper are independent of the detailed
modeling of enhanced mass transfer.

\subsubsection{General predictions}

\begin{table*}
\begin{center}
\caption[]{\label{t-res}Parameter and delays of calculated
outbursts. The obtained delays depend on where one measures them;
we give values at the beginning of the optical rise (index 0) and
at half the maximum optical flux (index 0.5). Here ``n'' denotes
the number of normal outbursts between a two superoutbursts and
$t_{\rm cycle}$ is the duration of a supercycle. The index ``rev''
labels the revised version of the TTIM (see
Sect.\,\ref{s-tti_rev}). The magnetic moment of the white dwarf
($\mu_{30}$) is given in units of $10^{30}$\,G\,cm$^3$ }
\setlength{\tabcolsep}{1.75ex}
\begin{tabular}{llllllllllr}
\hline
\hline
\noalign{\smallskip}
model &
$\dot{M}_{\mathrm{tr}}$ &
$\mu_{30}$ &
$\alpha_{\rm c}$ &
$\alpha_{\rm h}$ &
$t_{\rm cycle}$&
n &
\multicolumn{1}{c}{$\Delta_{\mathrm{EUV,0}}$}&
\multicolumn{1}{c}{$\Delta_{\mathrm{EUV,0.5}}$}&
\multicolumn{1}{c}{$\Delta_{\mathrm{UV,0}}$}&
\multicolumn{1}{c}{$\Delta_{\mathrm{UV,0.5}}$}
\\
&
$[10^{16}{\rm gs}^{-1}]$ &
&
&
& [d]
&
& [d]
& [d]
& [d]
& [d]\\
\noalign{\smallskip}\hline\noalign{\smallskip}
TTIM &  1 & 0 & 0.04 & 0.2 & 160 & 4 & 0.4 & 0.5 & 0.2 & 0.3 \\
TTIM$_{\rm rev}$ & 1 & 0 & 0.04 & 0.2 & 200 & 5 & 0.4 & 0.5 & 0.2 & 0.3\\
TTIM$_{\rm rev}$ & 2.5 & 0 & 0.04 & 0.2 & 130 & 4 & 0.4 & 0.5 & 0.2 & 0.3\\
EMTM & 1.25 & 0 & 0.04 & 0.2 & 180 & 5 & 0.4 & 0.5 & 0.2 & 0.3\\
EMTM & 1.62 & 0 & 0.04 & 0.2 & 120 & 3 & 0.4 & 0.5 & 0.2 & 0.3\\
EMTM & 1.5 & 0 & 0.045 & 0.195 & 200  & 7 & 0.6 & 0.5 & 0.4 & 0.3\\
EMTM$^*$ & 1.5 & 2 & 0.04 & 0.2 & 160  & 6 & 0.4 & 0.5 & 0.2 & 0.3\\
EMTM$^*$ & 1.85 & 2 & 0.05 & 0.185 & 100  & 4 & 0.6 & 0.7 & 0.4 & 0.5\\
EMTM & 1.25 & 2 & 0.04 & 0.2 & 170  & 4 & 0.3 & 0.5 & 0.2 & 0.3\\
\noalign{\smallskip}\hline\noalign{\smallskip}
\end{tabular}
\linebreak
\end{center}
 $^*$ Calculations performed with the parameters according to $\Mwd=0.86$ (see
Table\,\ref{t-par}). 
\end{table*}

\begin{figure}
\includegraphics[width=8.5cm, angle=0]{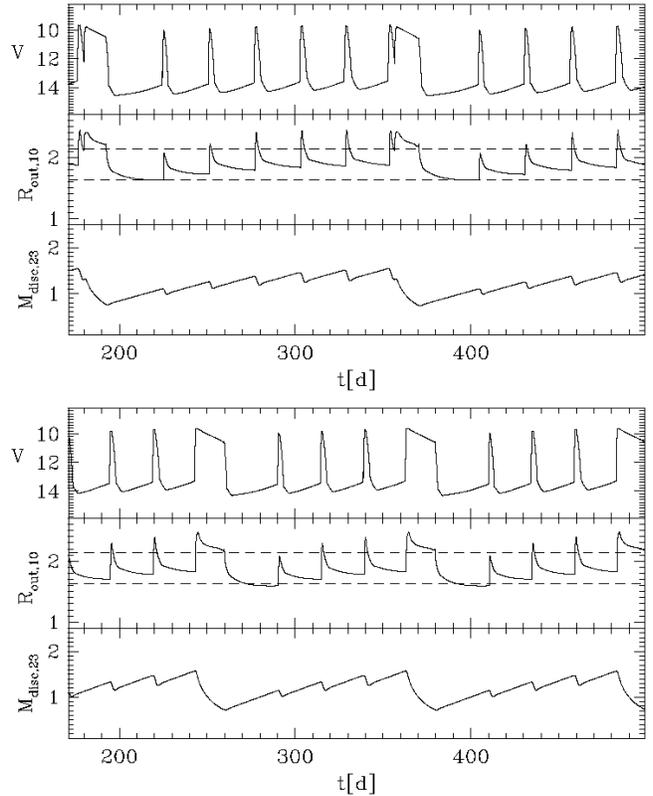}
\caption{\label{f-emt_mtr} The same as Fig.\,\ref{f-tti_mtr} but
for the EMTM. From top to bottom the panels show the evolution of
the visual magnitude, outer radius
($R_{\mathrm{out},10}=R_{\mathrm{out}}/10^{10}$\,cm), and the disc mass
($M_{\mathrm{disc,23}}=M_{\mathrm{disc}}/10^{23}$\,g). The mass
transfer rate is increased by $30\,\%$ in the lower panels. In the
EMTM a superoutburst is triggered when the disc mass reaches a
certain value while the size of the disc depends on the mean mass
transfer rate.}
\end{figure}

Fig.\,\ref{f-emt} shows the resulting light curves and the
evolution of \Rout\,. The model predicts five small outbursts
between the longer ones. As in the case of the revised TTIM all
outbursts except the rebrightening following the precursor are of
the inside--out type. The brightness difference between
superoutbursts and normal outbursts is smaller because the
mechanism which is causing the long superoutbursts (enhanced mass
transfer) is also present in normal outbursts.  The predicted
light curve clearly shows a precursor outburst which is much more
pronounced in the UV and EUV light curve than in the optical. The
optical - UV and optical - EUV time lags (defined by the rise
time) are identical to those obtained in the previous sections
(see Table\,\ref{t-res}). On the other hand, the predicted
superoutbursts are somewhat different: in the EMTM the decrease of
the optical and even more so the UV and EUV flux during
superoutburst is less drastic than in the TTIM.

The similarity of the obtained delays is
not surprising as they depend on the mass accretion rate after the
heating front has reached the
inner edge of the disc and on the viscous time scale. Both
quantities depend mainly on the values of $\alpha$ which we did
not change. The predicted precursor is caused by the fact that the
maximum accretion rates are delayed with respect to the optical
rise. Thus, when the heating front reached the outer edge of the
disc, the mass accretion rate (and hence also the mass transfer
rate) are still rather low so that a cooling front starts. At this
time, the mass accretion rate is still increasing. Once the maximum
mass transfer has been reached, an outside--in heating front which
overwhelms the precursor cooling front is triggered by an
EMT.

\begin{figure}
\includegraphics[width=6.5cm, angle=270]{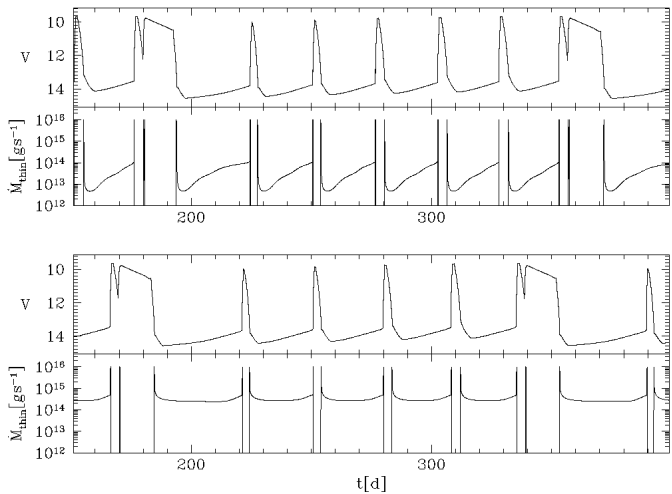}
\caption{\label{f-xevol} Optical light curves and mass accretion rates
when the inner region is assumed to emit X-rays ($\dot{M}_{\rm thin}$).
In the top panels we assumed the disc to extend down to the surface of the
white dwarf whereas the disc has been truncated during quiescence
in the bottom
panels. Apparently, in the latter case the accretion rate is higher and
almost constant during quiescence. X-ray flares are predicted
at the onset and end of the outbursts as well as at the end of pronounced
precursors.}
\end{figure}

\subsubsection{Dependence on the mass transfer rate}

In Fig.\,\ref{f-emt_mtr} we present light curves predicted by the
EMTM for different mass transfer rates. In contrast to
Sect.\,\ref{s-tti_mtr} we use only slightly different mass
transfer rates. The EUV and UV delays remained nearly unchanged
(see Table\,\ref{t-res}) and the predicted durations of the
superoutbursts do not change with the mass transfer rate. However,
increasing the value of the mean mass transfer rate ($\Mtr_0$)
significantly alters the predicted light curve: short outbursts
become less frequent and the duration of the supercycle is
significantly reduced. In addition, the precursor vanishes even
for slightly increased mean mass transfer rates
(Fig.\ref{f-emt_mtr}).

The reason for the completely different responses of the EMTM to
mass transfer variations lies in the triggering conditions for
superoutbursts. Whether an outbursts becomes a superoutburst or not is
independent of the outer radius reached during the expansion. It
depends on the maximum accretion rate during outburst which
increases with the mass of the disc. Hence, the irradiating flux
and the maximum mass transfer rate also depend on how much mass
has been stored in the disc. Increasing the mean mass transfer
rate therefore shortens the time between superoutbursts. Indeed,
as the mass accreted during superoutbursts remains constant\footnote{The
duration of superoutbursts depends on the strength
of enhanced mass transfer, i.e. $f_{\rm ill}$ (see
Eq.\,(\ref{eq-emt1}))} we find for the duration of a supercycle
\begin{equation}
t_{\rm SS}\propto\,\Mtr_0,
\end{equation}
when neglecting the mass accreted during normal outbursts. The
variations in the appearance of the precursor outburst are also
easy to understand considering the conditions causing a
superoutburst. As the mass of the disc remains unchanged, the disc
is smaller for higher mass transfer rates. The critical mass
accretion rate above which the disc is in the high state depends
sensitively on the radius (i.e.
$\dot{M}_{crit}^+\propto\,R^{2.68}$ \citep{hameuryetal98-1}). Thus
whether the disc stays long enough in the hot state to avoid a
precursor cooling front depends sensitively on $R_{\rm out}$. In
the case of high mass transfer, the hot disc expands slowly while
the accretion rate -- and hence the irradiating flux as well as the
mass transfer rate -- increase until the latter keeps the disc in the
hot state. For lower mass transfer rates, the disc is larger
during quiescence and the cooling front develops
shortly after the heating front has reached the
outer edge (before the enhanced mass
transfer will cause an outside--in heating front and the
rebrightening).

\subsection{Truncation of the inner disc}

\begin{figure}
\includegraphics[width=7.5cm, angle=0]{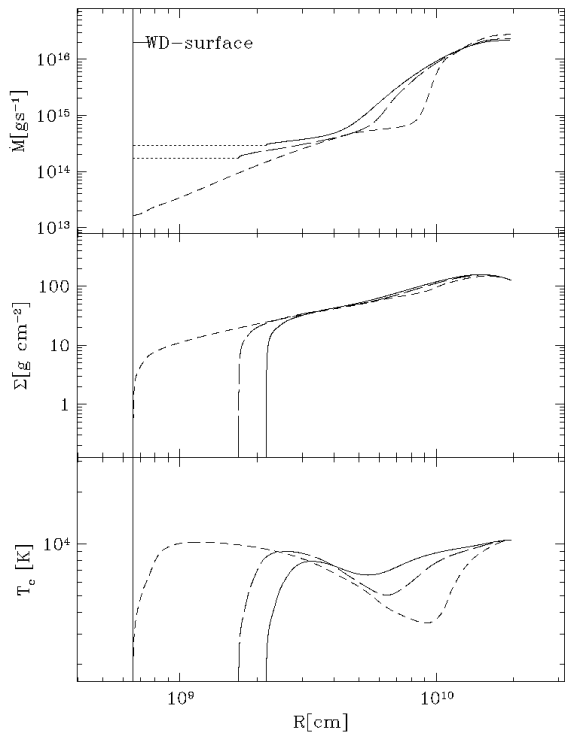}
\caption{\label{f-rad_tr}
Radial structures of accretion discs for $\mu_{30}=2$ (solid line),
1 (long dashed line), and without truncation ( $\mu_{30}=0$, dashed line).
For this set of calculations we assumed the higher value for the mass of the
white dwarf given in Table\,\ref{t-par}. The results, however, do not sensitively
depend  on the white dwarf mass (see Table\,\ref{t-res} and Sect.\,\ref{f-obsdel}).
The mass flow through the disc (top panel) is increasing, i.e.
$\partial\,\log\,\dot{M}/\partial \log\,R\sim\,2.3$. Thus the larger the
magnetospheric radius of the white dwarf, the larger is the supply of mass by
the disc to the inner (magnetically controlled) optically thin flow (dotted
lines). The other panels show the central
temperature (bottom) and the surface density (middle) as a function of
radius.
In all three cases
the disc is in very similar states during quiescence. The only
difference concerns the truncation radius which is essential only for the
mass accretion rate onto the white dwarf. }
\end{figure}

\begin{center}
\begin{figure}\begin{center}
\includegraphics[width=4.5cm, angle=270]{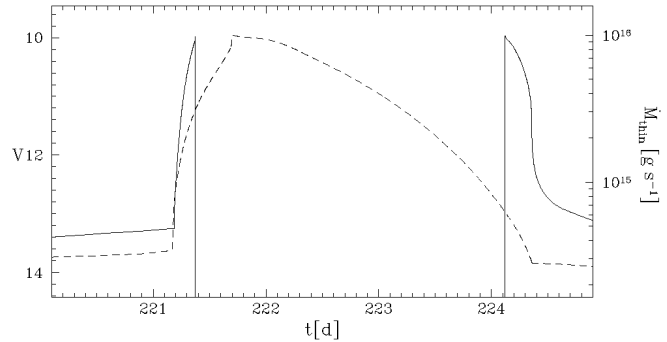}
\caption{\label{f-xrise} Optical outburst (dashed line) and the accretion rate
when the inner region is assumed to be optically thin (solid) representing the
X-ray emission. The inner disc
is assumed to be truncated during quiescence ($\mu_{30}=2$, see
Eq.\,(\ref{eq-rin})). During outburst the X-ray emission is suppressed
whereas the
calculations predict increased X-ray emission for $0.5-1$\,d at the beginning
and end of the outbursts.
}
\end{center}
\end{figure}
\end{center}

So far we concentrated on the predictions of the TTIM and the EMTM
in the UV and EUV wavelength range. As mentioned in
Sect.\,\ref{s-xrays} the observed X-ray emission is also very
important when discussing multi-wavelength properties of SU\,UMa
systems. The differences between the EMT and the TTI models
concern the outer disc regions during outburst when the accretion
rate is high and the boundary layer optically thick. The evolution
of the mass transfer rate below \Mcr (representing the X-ray light
curve) is therefore identical for both models. In contrast,
as shown in paper\,{\rm I}, truncation of the inner disc during
quiescence significantly affects the results. In this section we
present the evolution of the mass transfer rate when the boundary
layer is assumed to be optically thin for the EMT scenario and
focus on the role of truncation of the inner disc -- an important
ingredient of the DIM.

The idea of truncation arose when the significant discrepancy
between calculated accretions rates and those derived from X-ray
observations became apparent.
Indeed, in paper\,{\rm I} we showed that the
DIM agrees essentially better with the observed optical light
curve as well as the X-ray luminosity if the inner disc is
truncated.
During the last decade various mechanisms that can lead to disc
truncation have been proposed to bring into agreement
theory and observations (e.g. truncation by a magnetic field
\citep{lasotaetal95-1} and evaporation
\citep{meyer+meyer-hofmeister94-1}).
\citet{pandeletal03-1}
detected orbital modulations in the light curve of VW\,Hyi indicating 
the presence of a weak magnetic field.
However, analyzing in detail which mechanism truncates the inner disc is
beyond the scope of this paper. In the framework of the DIM
it is sufficient to assume the formation of an inner hole
during quiescence (see also Sect.\,4.2 of paper\,I).

In this section we present EMTM calculations including
truncation of the inner disc during quiescence by a weak magnetic
field:
\begin{equation}
R_{\mathrm{in}}= R_{\mathrm{M}}= 9.8\times10^8
\dot{M}_{15}^{-2/7}\Mwd^{-1/7}\mu_{30}^{4/7}\mathrm{cm}
\label{eq-rin}
\end{equation}
where $\mu_{30}$ is the magnetic moment of the white dwarf in
units of $10^{30}$\,G\,cm$^3$ \citep[see
e.g.][]{hameury+lasota02-1}.

Fig.\,\ref{f-xevol} compares the evolution of the accretion rate
when the boundary layer (or the hot inner flow) is assumed to be
optically thin. If the inner disc is truncated, the accretion rate
in this region is significantly higher and remains
nearly constant. This is easy to understand considering
the fact that the disc is {\em not} in a stationary state during
quiescence. Instead matter accumulates
and the mass flow rate in the disc increases with radius.
In Fig.\,\ref{f-rad_tr} we present the radial structure of the disc for
truncated and non-truncated discs using the higher estimate for the white
dwarf mass, i.e. $\Mwd=0.86\Msun$.
At first we consider the case without truncation
(dashed line in Fig.\,\ref{f-rad_tr}). We find a small accretion rate onto
the white dwarf ($\sim\,10^{13}$\,gs$^{-1}$) but a strong dependence of
the mass flow through the disc on the radius, i.e.
$\partial\,\log\,\dot{M}/\partial \log\,R\sim\,2.3$. (top panel of
Fig.\,\ref{f-rad_tr}).
Concerning truncated discs (long dashed and solid line in Fig.\,\ref{f-rad_tr}) the immediate
consequence of this result is: the larger the truncation radius the larger
is the mass supplied to the inner (X-ray emitting) flow by the disc.
The remaining outer disc is clearly not affected by
truncation during quiescence (middle and bottom panels of Fig.\,\ref{f-rad_tr}).

Once enough mass has been accumulated in the disc and the heating front
starts, the model predicts a sudden
increase of the X-ray emission. The magnetospheric radius decreases until the
disc has filled the inner hole and the boundary layer is assumed
to become optically thick, i.e. when $\Macc\geq\Mcr$ (see
Fig.\ref{f-xrise}). As we obtain only inside--out outbursts for the
standard form of the viscosity prescription, we predict that the
sudden X-ray flares at the onset of an optical outbursts are not
delayed with respect to the optical rise.
Because we assumed the inner region to
switch instantaneously between the optically thin and the optically thick
state when the accretion rate is passing \Mcr, we obtain also
increased X-ray emission at the end of pronounced precursor outbursts
(Fig.\,\ref{f-xevol}) which is probably not a very realistic prediction.

Concerning the UV and EUV delay we note that truncation slightly
shortens the delays at the onset of the outburst, i.e. by
$0.02$\,d (\UVon) and $0.03$\,d (\EUVon) respectively
as the heating
front is forced to start at a somewhat larger radius ($2.2\times10^9$cm
instead of $\sim1.1\times10^9$\,cm, see also paper\,I).

\begin{figure*}
\begin{center}
\includegraphics[width=9cm, angle=270]{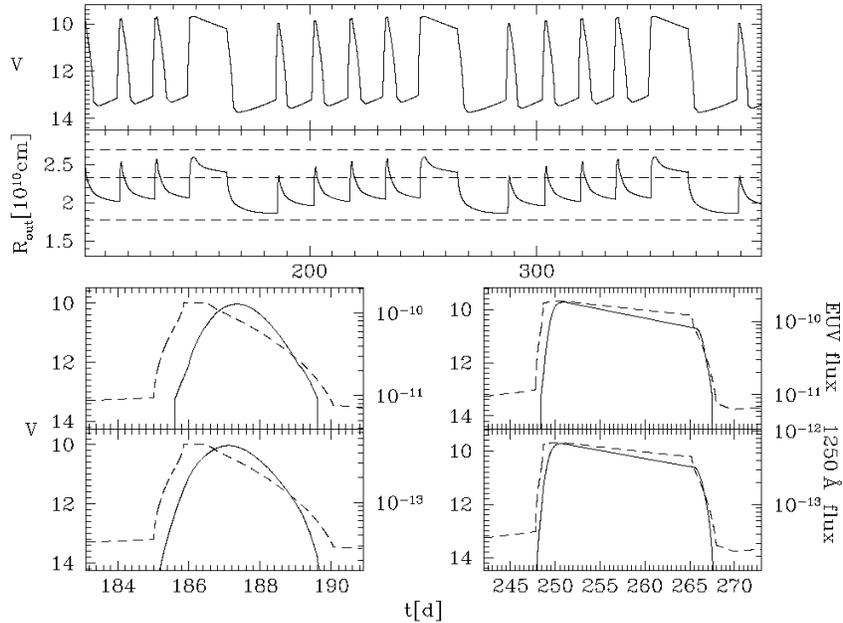}
\caption{\label{f-allow}Light curves calculated for a smaller ratio
of $\alh/\alc=0.185/0.5$ and the parameter according to $\Mwd=0.86\Msun$
(Table\,\ref{t-res}).
Due to the smaller ratio of $\alh/\alc$ we obtain longer delays in
very good agreement with the observations. In addition, the change of $\alpha$
leads to somewhat more frequent small outbursts.
The change of the system parameter due to the larger value assumed for
$\Mwd$ does not influence the general predictions of the model.
The only difference worth mentioning is the increase of the mean outer radius
of the disc. To produce light curves similar to the observed ones, this
requires to assume a higher value for $\Mtr$ and leads to somewhat longer
outbursts.}
\end{center}
\end{figure*}

\section{TTIM or EMTM: comparison with observations\label{s-sim_obs}}

It is well known that both models are able to produce a cycle of
several small outbursts between superoutbursts which more or less
resembles the observed one. In the previous sections of this paper
we additionally present multi-wavelength predictions of the TTIM
and the EMTM in order analyze the systematic features of the
models. Now we want to relate their characteristics to the
observed phenomena to see if we can decide which model has to
be preferred.

\subsection{Delays in normal outbursts \label{f-obsdel}}
Both models predict at the onset of the outbursts too short delays
($\UVon=0.2$,\,$\EUVon=0.4$\,days) when compared with the observed
ones ($\UVon\sim\,0.4$,\,$\EUVon\sim0.75$\,days). In earlier
publications it has been claimed that the outbursts in VW\,Hyi
must be of the outside--in type in order to explain the observed
long delays \citep[e.g.][]{smak98-1}. However, as shown in
paper\,$\rm{I}$, the delays calculated for outside--in outbursts
are even shorter than those of inside--out outbursts. Moreover,
using the TTIM or EMTM and the standard form for the viscosity, we
did not obtain outside--in normal outbursts even for
the highest mass transfer rates.

Nevertheless the discrepancy obtained between the
observed and predicted delays does {\em not} represent a
serious problem for the models. As mentioned in paper $\rm{I}$
(Sect.\,4.3), the delays measured during the early rise represent
the time it takes the disc to reach high accretion rates. After a
heating front has passed, the mass accretion rate of the disc
sensitively depends on $\alpha_{\rm h}/\alpha_{\rm  c}$. In the
calculations presented in Sect.\,\ref{s-sim} we always used
$\alpha_{\rm h}/\alpha_{\rm c}=0.2/0.04=5$.
To demonstrate the influence of the viscosity parameter on the predicted
delays we additionally performed simulations with reduced values of
$\alpha_{\rm h}/\alpha_{\rm c}$ (see Table\,\ref{t-res}).
In Fig.\,\ref{f-allow}
we show the multi-wavelength predictions of the EMTM for
$\alpha_{\rm c}/\alpha_{\rm h}=0.185/0.05=3.7$. Both \UVon\,
(0.4\,days) and \EUVon\,(0.6\,days) are significantly longer and
in better agreement with the observations. Hence a small
change of the parameter $\alpha$ can bring into agreement model
and observations. The change of $\alpha$ also leads to shorter quiescence
times as the viscous time scale during quiescence decreased.

The calculations presented in Fig.\,\ref{f-allow} are also calculated
with the larger mass of the white dwarf given in Table\,\ref{t-par}.
The increased mean outer radius $<R_{\mathrm{out}}>$ requires an increased
mass transfer rate, the superoutbursts are slightly longer (as a larger disc
contains more mass), and the tidal truncation radius as well as the 3:1
resonance radius are larger.
Nevertheless, the delays and the general predictions
of the model do not depend on the uncertainties related to the mass of the
white dwarf which becomes evident when
inspecting Table\,\ref{t-res}.

Concerning the EUV light curve we additionally wish to stress
the influence of the critical accretion rate above which the
boundary layer is assumed to become optically thick. Throughout
this paper we used the same value as in paper\,$\rm{I}$, i.e.
$\Mcr=10^{16}$g\,s$^{-1}$. This is justified for SS\,Cyg by the
observed hard X-ray emission. However, we have to be aware of the
fact that $\Mcr$ might be different in VW\,Hyi as it depends on
the mass and rotation of the white dwarf as well as on the value of
$\alpha$ assumed for the boundary layer
\citep{popham+narayan95-1}. The assumed value of $\Mcr$ is
essential for the predicted EUV delay at the onset of the
outburst: a higher value would increase $\EUVon$ and steepen the
predicted EUV light curve while a value significantly smaller than
$10^{16}$g\,s$^{-1}$ would cause essentially too short delays.

\subsection{X-rays and truncation}

The X-ray predictions of the TTIM and the EMTM are identical as
the differences of the models concern the outburst state and the
outer disc whereas X-ray emission is expected only during
quiescence, the very early optical rise and the late decline
(Figs.\,\ref{f-xevol},\,\ref{f-xrise}). Clearly, our approximation
of the X-ray emission agrees with the observed suppression of
X-rays during outburst.

Concerning the possible truncation of the inner disc during
quiescence we come to the same conclusion as in paper\,I: assuming
truncation leads to much more promising results as the predicted
accretion rate is equal to the one derived from observations
($3\times\,10^{14}$\,gs$^{-1}$) and remains almost constant during
quiescence (Fig.\,\ref{f-xevol}) as it is observed
\citep{hartmannetal99-1}. We want to underline here again that
this is because during quiescence the mass flow is increasing with
radius in the {\em non-stationary disc}. Therefore disc
truncation leads to an increased accretion rates in
the inner X-ray emitting region (Fig.\,\ref{f-rad_tr}).
As a final but important note on this subject we would like
to stress that good fits of quiescent dwarf-nova discs with
stationary disc models {\em contradict} the DIM.

Independent of truncation is the prediction of X-ray flares at the
beginning (simultaneously to the optical rise) and end of an
outburst. It has not yet been confronted with observations
but to fill this gap of observational evidence for the DIM is
certainly an interesting topic for future X-ray proposals.

\subsection{Superoutbursts}

We consider here the general characteristics of the predicted
superoutburst light curves. The precursor phenomenon will be
discussed in the next section.

Broadly speaking, the superoutbursts predicted by the TTIM and the
EMTM are quite similar.
The superoutburst brightness
relative to the normal outbursts
obtained with the EMTM is smaller than
that predicted by the TTIM. Both models predict during
superoutburst a decreasing flux at all wavelengths but this
effect is less drastic in the EMTM than in the TTIM (compare
Fig.\,\ref{f-tti_rev} and Fig.\,\ref{f-emt}). Although the details
of the above predictions depend on the arbitrary assumptions
related to the modeling of the tidal instability and the enhanced
mass transfer, these features are due to the fundamental
differences between both types of models:

\noindent (1) while the enhanced mass transfer also brightens
small outbursts the thermal-tidal instability appears only in
superoutbursts;

\noindent (2) whereas the TTI superoutburst is mainly fed by the
mass accumulated in the disc, the enhanced mass transfer provides
additional fuel during EMT superoutbursts.

We list the available observations concerning these differences
together with the values derived from our calculations in
parentheses:

\begin{enumerate}
\item superoutbursts are 0.5--1\,mag brighter than normal
outbursts (TTIM:$\sim\,1.5$; EMTM:$\sim\,0.5$) in the optical; at
UV wavelengths the maximum of normal and superoutbursts has been
measured with a single instrument only for an exceptionally faint
normal outburst \citep{polidan+holberg87-1}; and finally for the
EUV we can just mention the background subtracted deep survey EUVE
(DS) count rates given by \citet{maucheetal01-1} which are
somewhat larger (by a factor of $\sim\,1.5$) for superoutbursts
(EMTM:$\sim1.5$\,;TTIM:$\sim4$).

\item The measured decline during superoutburst is ranging from
$0.5$ to $1.5$\,mag (TTIM:$\sim\,2\,$mag; EMTM:\,$\sim\,1$\,mag))
in the optical; and by a factor of $\sim\,4\,$ at DS count
rates (EMTM:$\sim\,6$; TTIM:$\gappr\,10$).
\end{enumerate}

The EMTM seems to match the observations slightly better but,
because of the arbitrariness of the detailed modeling and the
involved simplifications we cannot exclude the TTIM so far.

\subsection{Precursors}

The TTIM as well as the EMTM predict precursor outbursts if
the enhanced tidal dissipation or respectively the
enhanced mass transfer are delayed by $\sim\,$1-2\,days with
respect to the optical rise of the superoutbursts. At short
wavelength, the predicted precursor outbursts are more pronounced
but in both models a cooling front is required to cause a
significant decrease of the EUV and UV emission. We
wish to stress that without the onset of a cooling front, both
models predict only a slow decrease at short wavelength and no
rebrightening (see Sect.\,\ref{s-tti_cls} about the standard
TTIM). Reproducing the precursor phenomenon requires a cooling
front. This is further confirmed by the observed local minima of
precursor outbursts which appear 4-5\,days after the initial
optical rise in very good agreement with our calculations (see
Fig.\,\ref{f-tti_rev} and \ref{f-emt}). In addition, the shape of
the observed UV precursor light curve presented by
\citet{polidan+holberg87-1} (see also Sect.\,\ref{s-uvprec}) is
identical to the UV light curve of a normal outburst just until
the re-brightening.

However, both models predicting a cooling front causes a
precursor at all wavelength; it is impossible to get a clear UV or
EUV precursor without at least a small decrease of the V--band
emission. The calculations we presented always show either pronounced
(1.5-2\,mag, Fig.\,\ref{f-tti_rev}) or no optical precursors
(Fig.\,\ref{f-emt_mtr}). Nevertheless, it is certainly possible to
construct light curves in which a UV and EUV precursor is clearly
visible although the predicted decrease of the optical emission
remains quite small. Concerning the EMTM we carried out this
fine-tuning by searching a value of $\Mtr_0$ close to the value
avoiding the precursor's cooling front. For
$\Mtr_0=1.563\times10^{16}$g\,s$^{-1}$ (within $5\%$ of the critical
value) we obtain a decrease of
$\sim\,1$\,mag in the optical light curve while the EUV  and UV
dips are essentially deeper and comparable to the observed ones.
Hence, according to the models every clear precursor at short
wavelengths should be accompanied by an optical decrease of $\sim\,1$\,mag.
So far the observed visual light curve
does not clearly constrain this prediction.

Although the predicted precursor outbursts are quite similar to
the observed ones, we should be aware that the detailed modeling is
rather arbitrary. Enhanced tidal dissipation as well as enhanced
mass transfer are based on {\em assumptions} which are reasonable
but artificially implemented in the standard DIM. In 1-D
simulations the tidal instability must be represented by a simple
parameterization and the response of a low mass star to
irradiation is still uncertain
(see Sect.\ref{s-prev}).
In view of this uncertainties it is rather
pointless to discuss whether the precursor outbursts predicted by
the revised TTIM or the EMTM resembles more the observed precursor
outbursts. Nevertheless, our numerical investigation provides a
decisive conclusion which is independent of the uncertainties
associated with modeling details: while the precursors predicted
by the TTIM are insensitive even to very {\em strong} mass
transfer variations the EMTM predicts that for even {\em small}
mass-transfer variations precursors might disappear. Considering
the variety of superoutbursts found in one single system, i.e.
VW\,Hyi \citep[see Sect.\,\ref{s-obs} and/or][]{bateson77-1}, this
represents a clear advantage of the EMTM.

\subsection{Optical light curves and variations of the mean mass transfer rate}

In dwarf novae the accretion disc always acts as a mass buffer and
it is difficult to directly derive the strength of mass transfer
variations from observed light curves. According to
state--of--the--art disc models the mass transfer rate must vary
dramatically in particular systems. The observed low states of
VY\,Scl stars \citep{hameury+lasota02-1} clearly indicate a
considerable decrease of the mass transfer rate, and in RX\,And
additional fluctuations are expected \citep{schreiberetal02-1}. If,
as in most dwarf novae rather regular outbursts are observed, only
small mass transfer variations are required. Indeed, assuming
variations of the mean mass transfer rate of $\sim\,30\%$

\noindent (i) allows to reproduce the change from normal to
anomalous as well as long to short outbursts in SS\,Cyg
(paper\,\rm{I});

\noindent (ii) is sufficient to explain Z\,Cam standstills
\citep{buat-menardetal01-2};

\noindent (iii) can within the EMTM account for the different
supercycle durations in single SU UMa systems e.g. VW\,Hyi;

\noindent (iv) and finally -- according again to the EMTM -- can
explain why precursors appear and vanish in the light curve of
VW\,Hyi.

Another argument for the EMTM concerns the SU\,UMa subclasses:
within the EMTM a dispersion of the mean mass transfer rates can
account for the different supercycle timescales in ER\,UMa and
ordinary SU\,UMa stars but one has to change the condition for the
tidal instability for different subclasses in the framework of the TTIM
\citep{buat-menard+hameury02-1,osaki95-1,osaki95-2}.

\section{Notes on earlier calculations\label{s-prev}}


\subsection{Delays}

\citet{smak98-1} analyzed the UV delay in dwarf novae and
concluded that the alleged problem of the UV delay is not a
problem if one uses the correct outer boundary condition and the
correct disc's size. For VW\,Hyi, Smak proposed outside--in
outbursts in order to bring into agreement the predictions of his
calculations and the observed delays. \citet{cannizzo01-1} comes to
the same conclusion when analyzing the EUV delay. We note the
following:

\noindent 1. Neither the UV nor the EUV delay are good indicators
for the ignition radius of the heating front (see also paper\,I);

\noindent 2. Both the TTI and EMT models predict outbursts of
the inside-out type except if a special form of viscosity and
an explicit radial dependence of $\alpha$ is assumed
\citep{ichikawa+osaki92-1}.

\noindent 3. The UV delay depends on the ratio of
$\alpha_c/\alpha_h$ while for the EUV delay the absolute value of
$\alpha_h$ (the viscous timescale in the hot state) is also
important.

\noindent Thus, in contradiction to earlier publications we
stress that outside--in outburst are not required to reproduce
the delays observed for VW\,Hyi.

\subsection{Irradiation of the secondary}

Finally we have to relate our results to the recent works of
\citet{osaki+meyer03-1} and \citet{smak04-1} discussing the possible role of
enhanced mass transfer during outbursts of SU\,UMa systems.

In their Sect.\,3 Osaki \& Meyer theoretically analyze the effects
of irradiation on the secondary and come to the conclusion that
the EMT scenario has to be abandoned because the $L_1$ point lies
deep in the shadow of the accretion disc and heat
transport from irradiated parts cannot be fast enough.
First Osaki \& Meyer claim
that Coriolis forces prevent hot matter to flow towards
$L_1$. In contrast they allege the formation of a so-called
``geostrophic'' flow, i.e. a flow at right angle to the driving
pressure gradient. Finally Osaki \& Meyer argue that diffusion as
the remaining mechanism transporting heat towards $L_1$ is much
too slow. Both results are incorrect because only the vicinity of
the $L_1$ point has been considered, where the normal component of
the rotation frequency is indeed not small. We note, however, that
even at small distance from $L_1$, this is no longer true and the
same holds for the arguments of Osaki \& Meyer.

Indeed \citet{smak04-1} presented a much more reliable approach
of calculating the effects
of irradiation of the secondary. He simplifies the problem by making the
reasonable assumption of isothermal layers inside and outside the
shadow zone and solves the equation of motion. He finds that
the enhanced mass transfer due to irradiation of the
secondary appears efficient in systems with short orbital
periods, i.e. $\Porb\lappr\,4$hr. Interestingly, the amount of
mass transfer enhancement derived by Smak for typical SU\,UMa
parameter (\Mtr/\Mtrn$\sim\,10-30$) is very similar to the one
assumed in our calculations. During outburst we obtain accretion
rates of \Macc$\sim\,5\times10^{17}$gs$^{-1}$. With an efficiency
of $f_{\mathrm{ill}}=0.3$ (Eq.\,(\ref{eq-emt2})) and
\Mtrn$=1.25\times10^{16}\,$gs$^{-1}$ this leads to a mass transfer
enhancement of \Mtr/\Mtrn$\sim\,12$. In addition, the values
derived from observations are also similar, e.g. \citet{vogt83-2}
derived \Mtr/\Mtrn$\sim\,15$ for VW\,Hyi.

To summarize the above, we conclude that irradiation induced
enhanced mass transfer is most likely present in SU\,UMa systems.

\section{Conclusion}

Using the disc instability model (DIM) and simple approximations
for the emission from the boundary layer we investigated the
systematics of the two scenarios proposed for SU\,UMa stars, i.e.
the tidal thermal instability (TTI) and the enhanced mass transfer
(EMT) models. We related the predictions of the model to
multi-wavelength observations of VW\,Hyi and SU\,UMa stars in
general. The results from this study are:

\begin{enumerate}

\item As shown previously, both models predict light curves
consisting of several small outbursts between longer and brighter
ones. Within the TTIM, long outbursts are triggered when the disc
reaches a critical outer radius ($\Rout=\Rto$) and the mass of the
disc at this time depends on the assumed mean mass transfer rate.
The EMTM predicts long outbursts when the disc mass reaches a
critical value. The outer radius of the disc at this very moment
depends on the assumed mass transfer rate during quiescence.

\item Using the standard prescription for the viscosity, both
models predict only inside--out outburst. The predicted UV and EUV
delays sensitively depend on $\alpha_c/\alpha_h$ and agree with
the observed ones for $\alpha_h/\alpha_c\lappr\,4$.

\item The calculated accretion rates during quiescence agree with
those derived from X-ray observations only if the inner disc is
truncated. The model also predicts X-ray flares at the beginning
and at the end of the outbursts similar to those observed for
SS\,Cyg. As we obtain for the EMTM as well as for the TTIM only
inside--out outbursts, both models predict no delay of the X-ray
flare with respect to the optical rise.

\item Both models predict precursor outbursts if the mechanism
which triggers the long outbursts, i.e. the tidal instability or
the enhanced mass transfer, is delayed with respect to the optical
rise. The predicted precursor outbursts are more pronounced at
shorter wavelength as observed. At all wavelength a precursor
outburst is obtained only if a cooling front has started and
every clear UV or EUV precursor should be accompanied by
a small ($\sim\,1$\,mag) decrease of visual emission.

\item The light curves predicted by the TTIM are rather
insensitive to even strong mass transfer variations. In contrast,
within the EMT scenario, the number of small outbursts, the
recurrence time of superoutbursts, and the appearance of precursor
outbursts sensitively depend on the mean mass transfer rate.
\end{enumerate}
Clearly, because of the approximative description of enhanced
mass transfer and the tidal instability we have not proven the
EMTM to be correct or the TTIM not to work. However, especially
because of point 5. the EMTM is currently the most promising
scenario: the observed variations in the optical light curve of VW
Hyi are naturally explained by mass transfer variations similar to
those predicted for Z\,Cam systems. In addition, within the EMTM
the different supercycle timescale of ordinary SU\,UMa stars
and ER\,UMa systems also has a simple explanation which we already know
from long orbital period dwarf novae: a dispersion of the mean
mass transfer rate at similar orbital periods.

\begin{acknowledgements}
MRS acknowledges funding by an individual Marie--Curie fellowship;
JMH and JPL were supported in part by the GDR PCHE. We thank Joe
Smak for sending us his article prior to publication and
we are gratefult to F. Bateson (VSS\,RASNZ) and J. Mattei (AAVOS) for
providing the long-term optical monitoring data of VW\,Hyi.
We acknowledge helpful comments and advice from Chris Mauche.
\end{acknowledgements}


\end{document}